\documentclass[12pt]{article}
\usepackage{mathtext}
\usepackage{graphicx}
\usepackage{amsfonts}
\usepackage{amssymb}
\usepackage{amsmath}
\newcommand{\s}{\sigma}
\newcommand{\la}{\lambda}

\begin{document}
\title{Quantum lock on dark states}

\author{Y.I.Ozhigov\thanks{ozhigov@cs.msu.su},\\
MSU of Lomonosov, VMK Faculty, SQI,\\
Institute of physics and technology RAS (FTIAN)}
\maketitle
\begin{abstract}
We propose quantum protection circuit (quantum lock), based on dark states of ensembles of two-level atoms in optical cavity. The secret key is the splitting of atoms into pairs, and publicly accessible part of the lock is the tensor product of EPR singlets, corresponding to the given splitting. To open the lock one must move synchronously pairs of atoms from the correct splitting to the other cavity; the lock will open if atoms do not emit photons. This scheme has perfect secrecy: it is impossible to hack it, even with effective solutions of any classical computational problems, in contrast to the RSA scheme. The method of obtaining dark states through Stark shift of atomic excitation energy is also proposed. This scheme makes possible to create secret keys of a few tens of atoms that is sufficient for the most practical applications.
\end{abstract}
\section{Introduction}

Data protection has several aspects. The most famous is called quantum cryptography, is associated with the distribution of secret key amongst authorized users of the system: Alice and Bob, which do not have, a-priori, any mutual information; rules of the game are unequal for them and all the others, unauthorized persons, including Eve (one of the first work is \cite{Wi}, see also \cite{BB} and many others). Quantum mechanics is able to significantly enhance the secrecy of cryptographic protocols: those for quantum cryptography have in the strict sense, absolute secrecy (see, for example, \cite{May}). 

Quantum information protection is not limited by key distribution. Its advantage over classical methods is known for many types of information processing. Quantum coin flipping (see \cite{Da}) and quantum bit commitment (see \cite{CK} and also \cite{Ma}) represent the cases when quantum methods help to defend the secret information. Contrariwise, the secure access can be effectively hacked by quantum computing (\cite{Sh}).

We consider the different problem: encryption of access and protection signatures. 
Here there is a publicly accessible part of the lock and the process of its actuation must lead to opening of the lock if and only if a key inserted by the user is correct. The most famous scheme of this type is RSA scheme (\cite{RSA}), in which the publicly accessible part of the lock is an integer $n$ and the secret key is the pair of its non-trivial divisors, $n=n_1n_2$. The secrecy of this, and similar schemes in that finding $n_1$ (it is enough to find one divisor) is a complex task, and the best classical algorithm for its solution requires the order of $exp((log(n))^{1/3})$ operations (\cite{L}). In 1994, P. Shor proposed quantum algorithm that solves the problem of factoring in the time, slightly exceeding the time $log^2(n)$ needed for multiplication of numbers themselves, hence the practical realization of a quantum computer would destroy the secrecy of RSA (\cite{Sh}).

We will consider a more general formulation of secrecy than algorithmic complexity. We consider as quickly solvable in principle (regardless of complexity) all classical problems, and assume that all processes and all physical devices used for protecting the information are available for analysis. For example, the work of the lock is publicly available. The secret, therefore, will be only the key itself in the form of a binary sequence. The secrecy is a basic principle of information protection (no back door, or peeping behind the cryptographer), hence the reliability of the whole process of information security should be based on the principles of quantum mechanics.

The naive attempt of quantum encryption would look as follows. The secret key $j$ is encrypted in the form of some publicly available quantum state $|\Psi_j\rangle$. Naturally, $|\Psi_j\rangle$ should not coincide with $|j\rangle$, since otherwise this lock opens with a simple measurement in the standard basis. If $|\Psi_j\rangle$ is a basic state, then the encryption is the usage of a classic, though unknown a-priori function, and therefore, in our conditions, does not provide absolute secrecy. If $|\Psi_j\rangle$ is some non-trivial superposition of the basis states, the operation of the lock is the measurement of $|\Psi_j\rangle$ in the certain basis that requires the storage of classical information about this non-trivial basis, which again, in our terms, means the absence of absolute secrecy. Thus, this primitive scheme of quantum encryption does not provide for absolute secrecy; quantum mechanics here, in fact, was not used.

We propose a scheme of encryption based on dark states of ensembles of two-level atoms. This scheme provides theoretically complete secrecy. Any attempt to open the lock by some allowed manipulation with the publicly accessible state $|\Psi\rangle$ without the knowledge of the key  results in loss of this state, whereas the lock will not open.

\section{Tavis-Cummings model}

To describe our scheme, we need to remind the general information about finite-dimensional model of quantum electrodynamics, proposed in the 60-s by E.Jaynes and F.Cummings in the work \cite{JC}, basing on the ideas of Dicke (\cite{D}). Their model reflects the interaction between the two level atom and one mode field in the optical cavity, and for the weak coupling constant it breaks down to two dimensional matrices corresponding to Rabi oscillations with the different enregies. This model was generalized by M.Tavis in \cite{T}, who investigated the ensembles of many two level atoms in the cavity interacting with each other through the one mode field only; we call this model Tavis-Cummings model (TC).

Hamiltonian of TC model for the weak interaction $g\ll h\omega$ (RWA approximation) looks as follows
\begin{equation}
H_{TC}=h\omega a^+a+h\omega\sum\limits_i\s_i^+\s_i+\sum\limits_ig_i(a^+\s_i+a\s_i^+)
\label{TC}
\end{equation}
where $\omega$ is the frequency of photon, which can live in the cavity in the time enough for our aims; if the half of its wave length is a divider of the lenght $L$ of cavity, this time (theoretically) can be unlimited. We suppose that the frequency of the atomic excitation equals $\omega$ that guarantees the interaction of the cavity photon with the atom by emission-absorption, $g_i$ is the intensity of interaction between the field and $i$-th atom, $a^{(+)}$ are operators of photon annihilation (creation), $\s_i^{(+)}$ are operators of annihilation (creation) of the $i$-th atom excitation. The intensity of interaction depends on some parameters: $g_i=\sqrt{h\omega/V}d_aE(x_i)$, where $V$ is the effective volume of the cavity, $d_a$- dipole momentum of each atom, $E(x_i)=sin(\pi x_i/L)$ - the spatial coordinate of $i$-th atom along the main axis of the cavity, $L$ is the length of cavity. 

Dark states of the atomic ensemble is such a state $|\Psi\rangle_{at}$, in which it cannot emit a photon, that is the set of dark states is $Ker({\bar\sigma})$, where $\bar\sigma=\sum\limits_{i=1}^ng_i\sigma_i$.

The example of dark states is the so called singlet states, which have the form 
\begin{equation}
\label{singlets}
|\Psi_K\rangle_{at}=\bigotimes\limits_{(i,j)\in K}(\gamma_j|0_i1_j\rangle-\gamma_i|1_i0_j\rangle)
\end{equation}
where $K$ is some splitting of the set of all qubits to pairs: $(1,j_i),(2,j_2),...,(n,j_n)$ for the different $j_1,j_2,...,j_n$, and we use the notation: $\gamma_i=g_1g_2...g_n/g_i,\ i=1,2,...,n$. We can also supplement a singlet state by multipliers of the form $|0\rangle$ - it will remain dark. The form \eqref{singlets} is not symmetric because the location of atoms in the cavity may be different. We can slowly move them using optical tweezers so that their coordinates $x_i$ becomes equal (but they will remain at the distance excuding direct dipole-dipole interaction); if the movement is slow enough, it does not destroy darkness by adiabatic theorem and we than obtain each EPR singlet in the symmetric form $|0_i1_j\rangle-|1_i0_j\rangle$. Factually, we will apply this movement only for one pair of atoms and there always will be the place inside the cavity for such a movement.

Each singlet in the state \eqref{singlets} is dark, because the attempt to emit a photon from one of two atoms is blocked by the similar attempt from the other - with the same absolute value of amplitude and the different sign. It is also evident that in the symmetric case, when all $g_i$ are equal, the states \eqref{singlets} caannot absorb a photon. It means that the states \eqref{singlets} do not interact with light and the subspace spanned by them is decoherence free, because the interaction between light and matter is the main source of decoherence. Such states are eigenstates of Tavis Cummings Hamiltonian and we will use such a state as the publicly available part of the quantum lock. 

\section{Creation and annihilation of singlet states}

Building of the lock and its work requires the creation of singlet states that we can do by Stark effect in the atoms in cavity. 

At first we consider the preparation of singlet states in the optical cavity by Stark effect.
To prepare one EPR singlet we start from the state of the form $|\psi (0)\rangle=|1\rangle_p|00\rangle_a$ with one photon and two atoms in ground states; where the frequency of the first atom is shifted by Stark effect, which we ensure by the proper voltage. To do this we must place atoms in the different locations along the central axis of cavity so that their coupling constants with the field will be different. This means that atomic ground states $|0\rangle_a^1$ and $|0\rangle_a^2$ have the different senses: it differ by location, by coefficients $g_1, g_2$, and by frequencies $\omega_a^1, \ \omega_a^2$ of transitions $|0\rangle\leftrightarrow |1\rangle$. 

We call S jump the following process of the evolution $|\psi (t)\rangle$ (see Figure 1). We wait some time $\Delta t$, which is choosen at random, and then abruptly turn off the voltage, which leads to Stark shift of the frequence of excitation for one single atom in the potential. Let  frequencies and coupling constant for our atoms without S jump look as $\omega, \omega, g_1,g_2$, and S jump applied to the first atom results in the following values for these parameters: $\omega +ds, \omega, g_1+dg,g_2$, e.g. it slightly changes the parameters of the first atom. 
After S jump we obtain EPR singlet on our two atoms if no photon is detected; this happens with some nonzero probability. 

It has been shown that the outcome of EPR singlets varies depending on the frequency shift value and can reach about $0.01\%$ with the available shift value (see \cite{Oz}) - this evaluation is rather modest than optimal; in all cases the proposed scheme of dark state fabrication is certainly realistic.

To prepare a singlet state of the form \eqref{singlets} we repeat the described prosedure with the next pair of atoms, etc. Given a secret key - a split $K$ of all atoms to pairs we thus obtain the singlet state $|\Psi_K\rangle$, corresponding to $K$.

\begin{figure}
\centering
\includegraphics[width=1.0\textwidth]{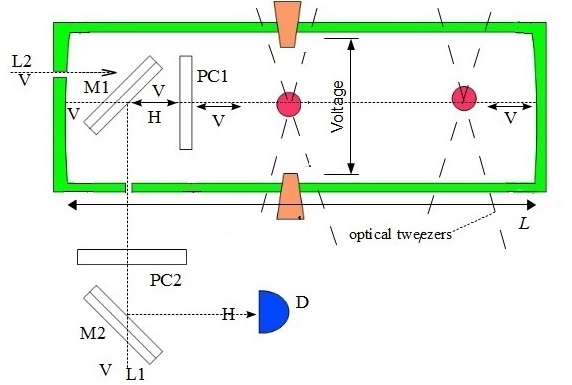}
 \caption{Preparation of dark singlet in optical cavity. Initially, Pockels cells PC1 and PC2 are both switched on. Photon flies from Laser L1 with vertical polarization $V$, changes polarization to $H$ after PC2, reflects from the mirror M1 and change its polarization to $V$ after PC1. Then PC1 must be switched off before photon comes back reflecting from the right wall of cavity. Photon will then locked inside the cavity. After arbitrary time frame PC1 switches on again and photon change the $V$ polarization back to $H$, passes through PC2, which is switched off, reflects from M2 and comes to detector $D$. Alternative way: photon comes from laser L2 with $V$ polarization and becomes locked in the cavity, then PC1 switches on, and photon comes to detector; PC2 is not needed. Figure is taken from the paper \cite{Oz}.}  

\end{figure}

\section{The work of quantum lock}

The quantum lock consists of two resonators connected by a path, through which we can move atoms from one cavity to the other, using optical tweezers. Each resonator has the Pockels cell that reflects photons emerging in the cavity to the control detectors adjacent to it (see Figure 2).

The working state of the lock is the state $|\Psi_K\rangle$ of atomic ensemble in the main cavity. The user types the password, which is some partition $K'$ of the ensemble to the pairs $a_i,b_i,\ i=1,2,...,n$, $a_i,b_i\in B$. The verification looks as follows. All pairs $a_i,b_j$ in turn are moved slowly and synchronously from the main cavity to the control cavity. The password is accepted if and only if at each step the both two controlling detectors keep silence during the movement of atoms and after several switchings on the shift potential ($S$- jumps) in the controlling cavity the detector D2 clicks. It happens exactly in the case $K=K'$. 

Indeed, if the  chosen pair $a_i,b_i$ matches the partition $K$, the current atomic state has the form 
$|\Psi_0\rangle\otimes |s_{a_i,b_j}\rangle$, and when moving the pair $a_i,b_i$ from the main cavity to the controlling this pair will remain in the singlet state and cannot emit a photon due to the synchronous movement, because the potential photons, which could arise by each atom are identical and their emissions are thus blocked by each other. On the other hand, if the pair $a_i,b_i$ does not match $K$, there exist two other pairs in $K$ of the form $a_i,b'_i$ and $a'_i,b_ia$. When moving the pair $a_i,b_i$ between cavities then the photons, which atoms $a_i$ and $a'_i$ can emit as well as atoms $a'_i$ and $b_i$ will not be identical that means the possibility of the equiprobable emission of a photon by each of these 4 atoms. In particular, with the probability each from $a_i$-th and $a'_i$-th atoms will emit the photon with probability 1/2, and the same for $a'_i$-th and $b_i$-th atoms, and all these photons will be reflected by the corresponding Pockels cell and reach the corresponding detector. Hence, in each case either D1 or D2 clicks. Here we assume that the detectors are ideal. Practically, with the non ideal detectors, we have only take more atoms in the ensemble: the probability to guess the main part of the password will be negligible and on one wrong choice of the pair $a_i,b_i$ one of the detectors will click.

\begin{figure}
\begin{center}
\includegraphics[height=0.6\textwidth]{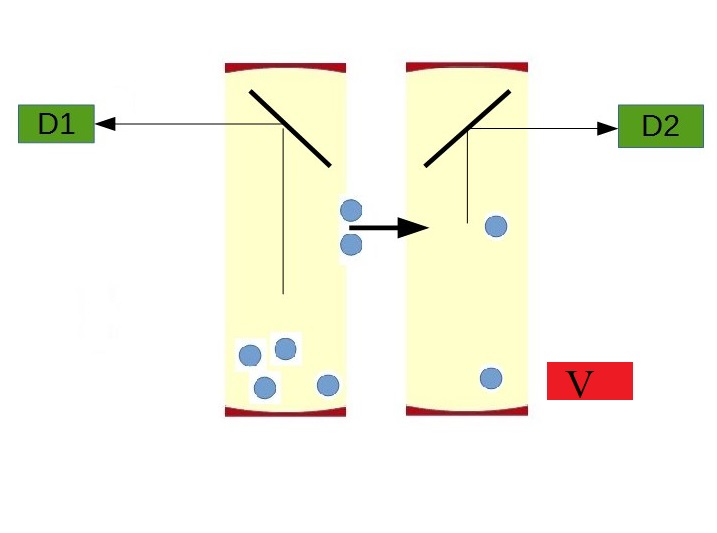}
\caption{Verification of the password for the quantum lock. The sequential pair is moving to the controlling cavity. One from the detectors must click if this pair does not match to the right password. }
\label{fig:nps}
\end{center}
\end{figure}

In the case when the pair $a_i,b_i$ is wrong, the probability of the photon emission by atoms $a_i$ (or $b_i$) is 1/2. If they emit a photon in the time when they are between cavities, this photon will not be detected; but if the emission takes place inside the controlling cavity, it hits detector $D2$. In all cases when the pair $a_i,b_i$ is wrong a photon appears; it passes by detectors only if it was emitted between cavities. In the last case the fact of emission can be checked subjecting the pair trapped in the controlling cavity by the action of non uniform potential causing Stark shift of atomic frequency that shows the excitation of one of atoms $a_i,b_i$; if these atoms are in the groundstate the current pair is considered as wrong.  

The proposed scheme of quantum lock does not require increasing the key length for privacy amplification beyond the value, which is necessary to minimize the probability of simply guessing the key: incorrect password value, with perfect lock operation is detected with absolute certainty. The length of the password, thus, in the ideal case, may be about 24 atoms, which will ensure that the probability of accidental guessing the password is not more than one hundred millionth that is an  acceptable level of secrecy for most practical applications.

The weak point of the proposed scheme is necessity to move atoms synchronously; if we let the asynchronous movement the probability arises that we consider the correct pair $a_i,b_i$ as wrong, because the photon, emitted by our pair due asynchronous movement can fly outside cavities and the detector D2 will not click after final S jumps. This situation is typical for classical locks as well; perghaps, to minimize its probability the longer key can be needed.

\section{Conclusion}

We have proposed the scheme of quantum password verification in which password is the split of two level atomic ensemble in the optical cavity, and the open part of the lock is the singlet-like quantum state, corresponding to the secret password. The verification of this password requires moving of atomic pairs from one cavity to the other and application of Stark shift resulted from the external potential. This scheme has the potential absolute security: it is impossible to crack it even all classical information (but the secret password itself) about the lock is known. This is the advantage of quantum lock on singlets over classical security methods. We have also shown how to prepare singlet states for the quantum lock: it can be done by non uniform potential, which makes Stark shift on the frequency of one selected atom from EPR pair; this method gives the portion of singlets sufficient for building the secret key of about 25-30 atoms that guarantees secrecy for most practical applications. 

\section{Acknowledgements}
The work is supported by Russian Foundation for Basic Research, grant 15-01-06132 a.

\end{document}